\documentclass[conference,letterpaper]{IEEEtran}

\usepackage[utf8]{inputenc} 
\usepackage[T1]{fontenc}
\usepackage{url}
\usepackage{ifthen}
\usepackage{cite}
\usepackage[cmex10]{amsmath} % Use the [cmex10] option to ensure complicance
\usepackage{graphicx}
\usepackage{caption}
\usepackage{xcolor}

\begin{document}
\title{Understanding the Error Sensitivity of Privacy-Aware Computing}

%%% Several authors with up to three affiliations:
\author{
    \IEEEauthorblockN{Matías Mazzanti\textsuperscript{1}, Esteban Mocskos\textsuperscript{1}, Augusto Vega\textsuperscript{2}, Pradip Bose\textsuperscript{2}\\}
    \IEEEauthorblockA{
        \textit{\textsuperscript{1}University of Buenos Aires (Argentina), \textsuperscript{2}IBM T. J. Watson Research Center (NY, USA)}}
}
%\author{%
%  \IEEEauthorblockN{Author 1}
%  \IEEEauthorblockA{Department of Electrical Engineering \\
%                    University 1\\
%                    City 1\\
%                    Email: author1@university1.edu}
%  \and
%  \IEEEauthorblockN{Author 2 and Author 3}
%  \IEEEauthorblockA{Research Center XY\\ 
%                    City 2\\
%                    Email: \{author2, author3\}@research-center.com}
%}

\maketitle

\begin{abstract}
    Homomorphic Encryption (HE) enables secure computation on encrypted data without decryption, allowing a great opportunity for privacy-preserving computation. In particular, domains such as healthcare, finance, and government, where data privacy and security are of utmost importance, can benefit from HE by enabling third-party computation and services on sensitive data. In other words, HE constitutes the ``Holy Grail'' of cryptography: data remains encrypted \textit{all the time}, being protected while in use.

    HE's security guarantees rely on noise added to data to make relatively simple problems computationally intractable. This error-centric intrinsic HE mechanism generates new challenges related to the fault tolerance and robustness of HE itself: hardware- and software-induced errors during HE operation can easily evade traditional error detection and correction mechanisms, resulting in silent data corruption (SDC).

    In this work, we motivate a thorough discussion regarding the sensitivity of HE applications to bit faults and provide a detailed error characterization study of CKKS (Cheon-Kim-Kim-Song). This is one of the most popular HE schemes due to its fixed-point arithmetic support for AI and machine learning applications. We also delve into the impact of the residue number system (RNS) and the number theoretic transform (NTT), two widely adopted HE optimization techniques, on CKKS' error sensitivity. To the best of our knowledge, this is the first work that looks into the robustness and error sensitivity of homomorphic encryption and, as such, it can pave the way for critical future work in this area.
\end{abstract}

\section{Introduction and Background}
\label{sec:introduction}

\begin{figure}[!ht]
  \centering
    \includegraphics[width=0.95\columnwidth]{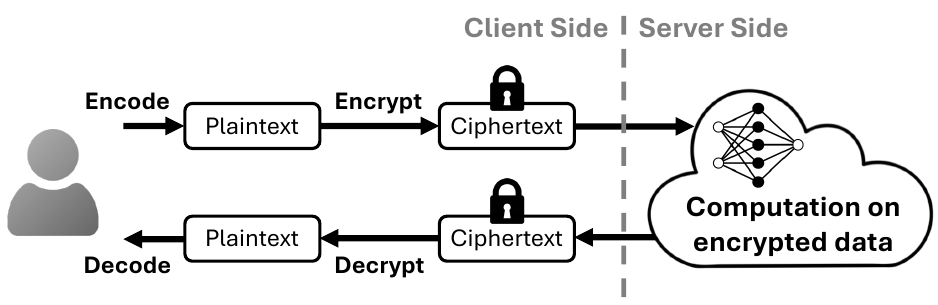}
	\caption{Homomorphic encryption used in a typical business application: users send their data encrypted to a third-party service provider, where data is processed in its encrypted form.}
	\label{fig:he_overview}
\end{figure}

Cloud computing has profoundly changed the way businesses and individuals use, process, and manage their data \cite{Konstantinos2015}. Despite the benefits of this approach, its adoption exacerbates some existing problems and generates new ones related to the security and privacy of user data~\cite{Sen2015}. In particular, delivering data to a third party to be processed opens up different possibilities of compromising them, both by improper access or by a decision of the provider to give access to the data without authorization.
One of the solutions to this problem resides in the use of data encryption schemes, which allow the data to only be interpreted by the holder of the key that allows decryption~\cite{Williams1980,Elgamal1986}.
However, most of the cryptography schemes do not allow computing directly on the encrypted data; it is necessary to decrypt before processing them. 
In this way, an external cloud computing provider can access the unencrypted data and, therefore, be able to compromise them. 
\textit{Homomorphic Encryption} (HE) solves this challenge by allowing \textbf{computation on encrypted data}~\cite{Yi2014}. 
Although HE schemes have high computational and memory requirements, which have limited so far their widespread adoption, the great interest in developing new schemes and the recent development of aggressive optimizations (at both algorithm and hardware levels) has opened the door to its use in real-world applications. Figure~\ref{fig:he_overview} presents an overview of a typical HE setting that involves a client that makes use of third-party cloud services.

Multiple HE schemes exist today, with different characteristics and supported capabilities. In this work, we focus on CKKS (Cheon-Kim-Kim-Song)~\cite{Cheon2017,Cheon2019}, a popular HE scheme for AI and machine learning applications due to its fixed-point arithmetic support. Fundamentally, schemes like CKKS base their hardness on a simple idea: add small errors (``noise'') to data to make relatively simple problems computationally intractable, an approach known as \textit{Learning With Errors} (LWE)~\cite{10.1145/1060590.1060603}. In other words, HE operations (like additions and multiplications) take place within a ``noisy'' domain.

As HE operates with noisy (or ``consciously erroneous'') data, a consequent question arises: \textit{how can we distinguish between HE's deliberately introduced error and error resulting from faulty hardware or software?} The answer, although not straightforward, motivates us to conduct a thorough study of the sensitivity of CKKS to bit faults (``flips''), resulting in a detailed error characterization study of this scheme.

\subsection{Silent Data Corruption}

\begin{figure}[!ht]
  \centering
    \includegraphics[width=0.9\columnwidth]{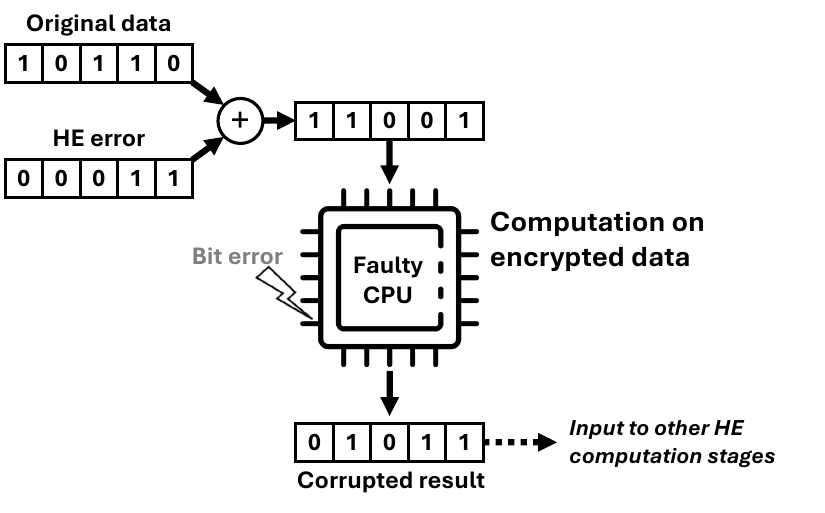}
	\caption{Illustrative scenario of a SDC case induced by a faulty CPU. The corrupted result incorporates both the HE error and the faulty hardware error.}
	\label{fig:he_error_example}
\end{figure}

Errors across CKKS stages (encoding, encryption, decryption, and decoding) can result in two scenarios: HE operation ``breaks''\footnote{The FHE library used in this work (OpenFHE) detects the data alteration and finishes its execution with an assertion error.} and the error is detected, or HE operation does not break and the error propagates across stages and ends up on \textit{silent data corruption} (SDC). The latter is the dangerous case and, as it has been widely studied and reported, hardware- and software-induced SDC happens, even in today's cutting edge systems and large-scale datacenters~\cite{dixit2021silentdatacorruptionsscale}. The nasty aspect about hardware and software errors in HE applications emerges from the very same error-centric intrinsic operation of HE schemes. Once in the HE domain, data becomes noisy by construction and, if additional error occurs due to faulty hardware or software, such error camouflages within the HE error and becomes very hard to detect. Figure~\ref{fig:he_error_example} presents a cartoonish illustration of this idea, where original data (plaintext) is ``encrypted'' by adding random HE error (noise)\footnote{In practice, encryption involves additional steps not shown in the figure.}.

\section{Error Resilience Analysis}
\label{sec:error_resilience_analysis}

\begin{figure}[!ht]
  \centering
    \includegraphics[width=1\columnwidth]{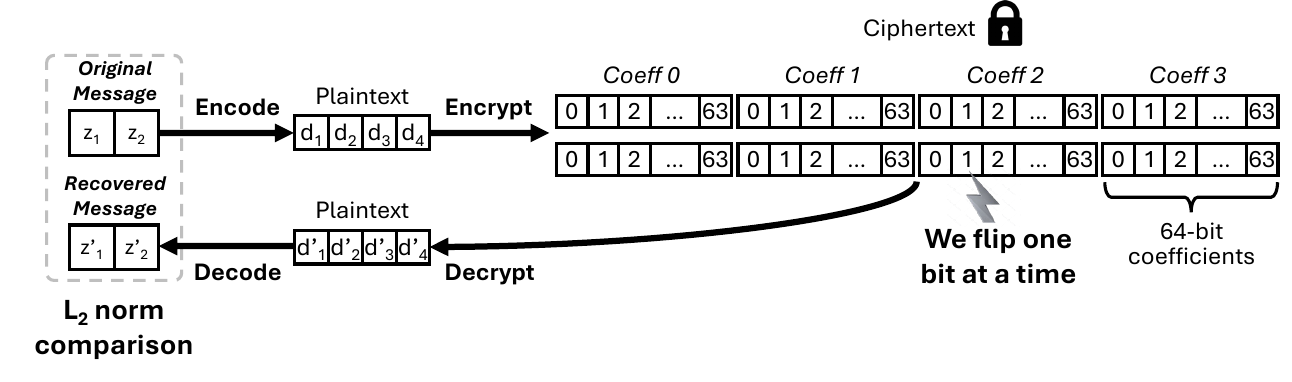}
	\caption{Error injection methodology.}
	\label{fig:methodology}
\end{figure}

This section presents preliminary error sensitivity results for CKKS. As illustrated in Figure~\ref{fig:he_overview}, the \textit{encoding} stage transforms user's input data into a \textit{plaintext}, a polynomial of degree $N$ that we will call $p(X)$. This plaintext is then encrypted into a \textit{ciphertext}, a pair of polynomials of degree $N$ each, that we will call $c = (c_0(X), c_1(X))$. We adopt a single-bit error model. As such, each bit of every polynomial coefficient (in both the plaintext and the ciphertext) is flipped in sequence for the set of single-bit-flip fault injection experiments. After each bit error injection, we execute the entire HE pipeline, and compare the recovered data after decoding (last stage) against the original data. This methodology is depicted in Figure~\ref{fig:methodology}, where two input elements (original message) are encoded into a 4-element polynomial (plaintext) and encrypted into two 4-element polynomials (ciphertext). A coefficient bit is flipped at a time before decryption and decoding. The recovered message is compared against the original one using the $L_2$ norm. We execute all runs on an Intel i7-11700 CPU with 32 GB RAM and Arch Linux 257.5-1, using the CKKS implementation from the OpenFHE library~\cite{OpenFHE}, which includes native RNS and NTT support and 64-bit coefficient representation.

A first observation is that the error behavior is similar when bit errors occur in coefficients of  $p(X)$ (the plaintext) and in coefficients of $c_0(X)$, one of the two polynomials in $c$ (the ciphertext). For this reason, and due to space constraints, in this paper we focus our campaign on errors in $c = (c_0(X), c_1(X))$.

This initial study emphasizes the occurrence of bit errors specifically during the encoding/encryption (and corresponding decoding/decryption) stages. It does not address errors that may arise during computations performed on the encrypted data, which will be the subject of future research.

\subsection{Error Position Impact}

\begin{figure}[!ht]
    \centering
    \includegraphics[width=0.8\linewidth]{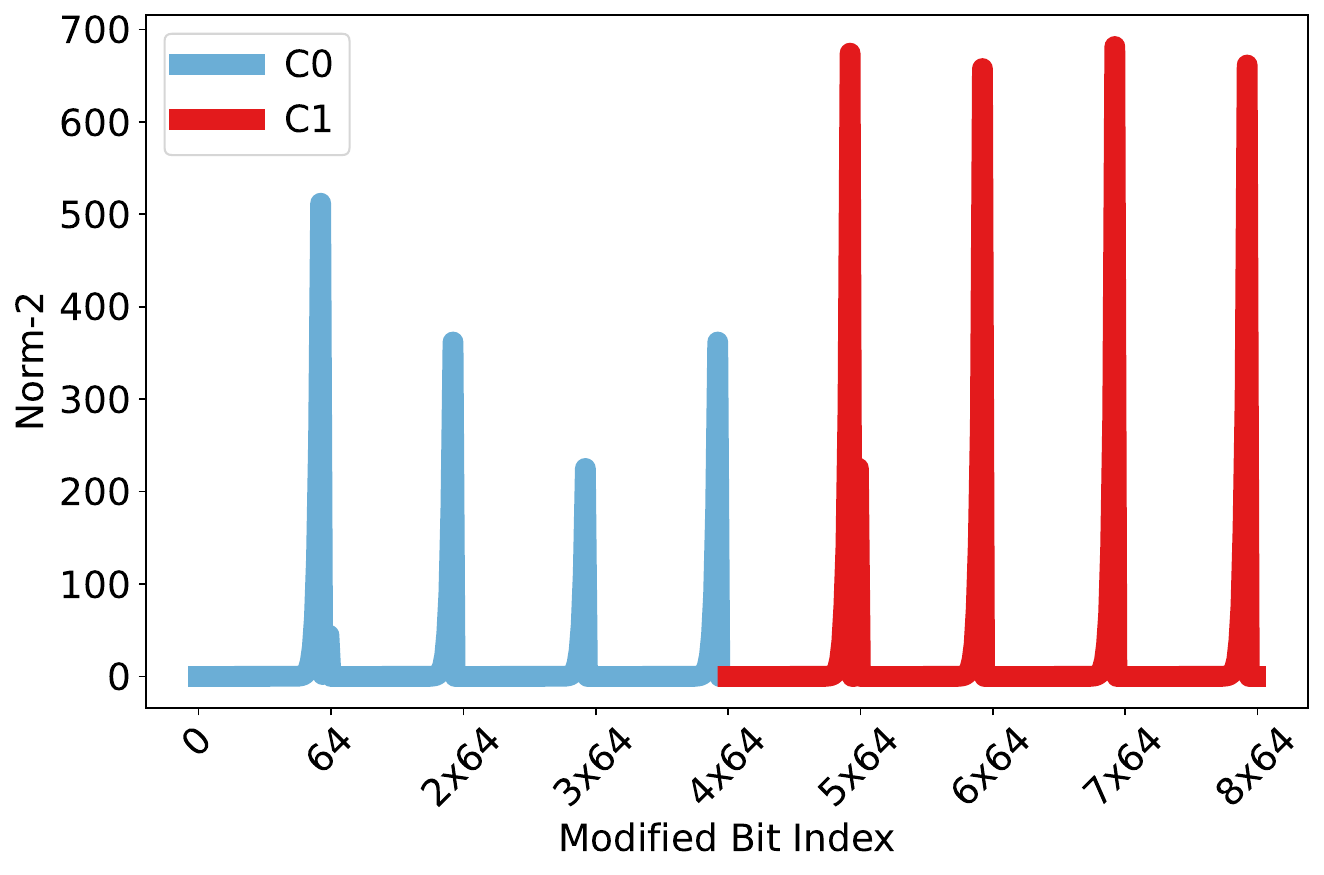}
    \caption{$L_2$ norm error of a single-bit flip in the ciphertext. The X-axis shows the position of the modified bit. Polynomial degree $N=4$.} 
    \label{fig:error-on-encryption}
\end{figure}

This section examines the impact of a single bit error introduced in the ciphertext $c = (c_0(X), c_1(X))$. We consider a simple scenario where both polynomials $c_0(X)$ and $c_1(X)$ consist of four 64-bit coefficients each. Figure~\ref{fig:error-on-encryption} illustrates the $L_2$ norm error of the output after decoding, in comparison to the original input. A somewhat expected observation is that the magnitude of the error increases with the importance of the altered bit within each coefficient. In particular, modifications to the first 50 bits of each coefficient result in negligible effects on the recovered output. Additionally, we observe that $c_1(X)$ exhibits more pronounced error peaks than $c_0(X)$. This phenomenon can be understood by examining the CKKS decryption process (Equation~\ref{eq:decryption}), where polynomial $c_1(X)$ is multiplied by the secret key $s$ (another polynomial), leading to the dispersion of the error across the coefficients.

\begin{equation}\label{eq:decryption}
    m' = [c_0 + c_1\times s]_Q
\end{equation}

\subsection{Scale Factor ($\Delta$) Impact}

\begin{figure}[!ht]
  \centering
    \includegraphics[width=0.9\linewidth]{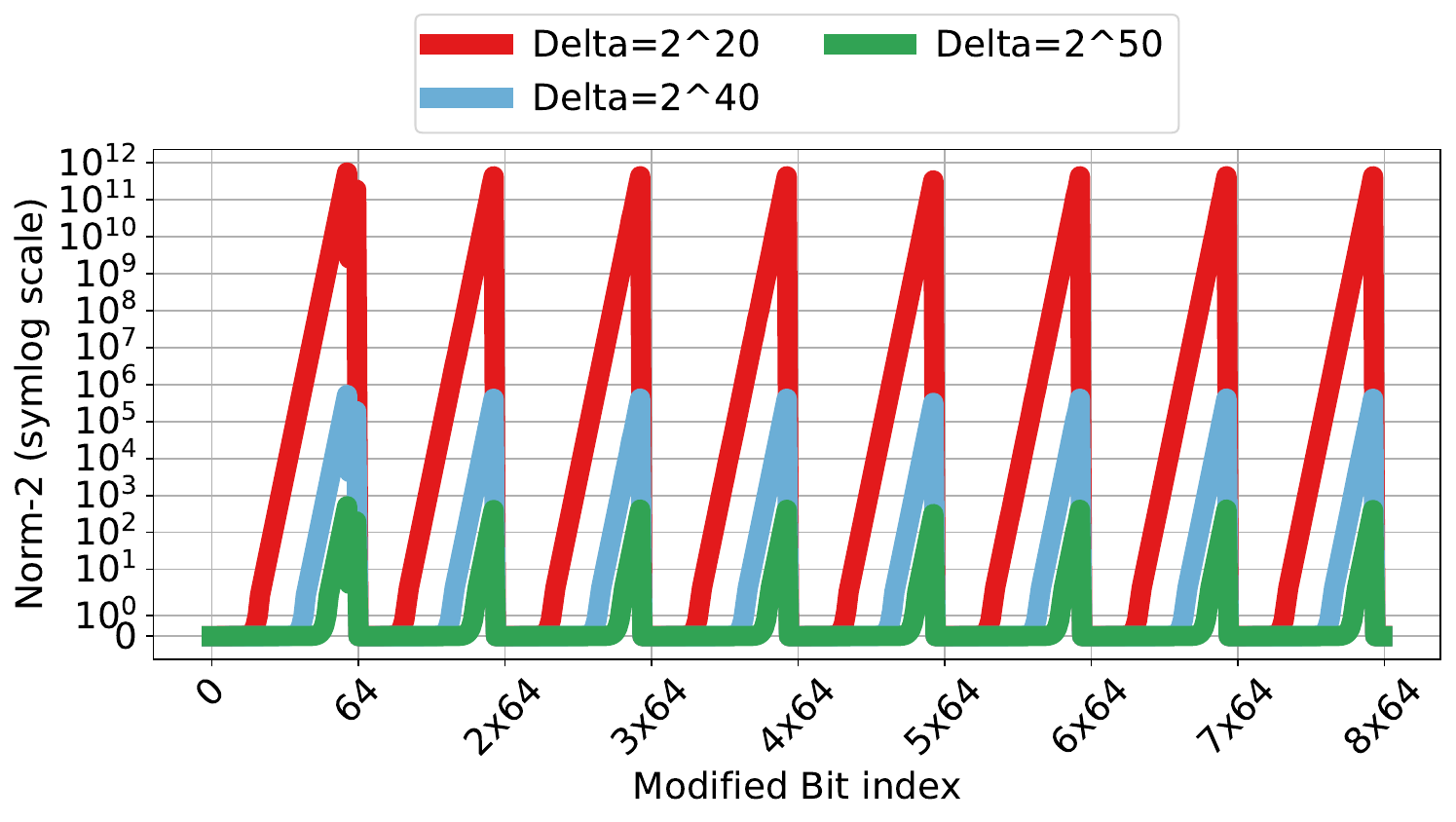}
    \caption{$L_2$ norm error of a single-bit flip in the ciphertext using different scale factors $\Delta$. The X-axis shows the position of the modified bit. Polynomial degree $N=4$.}
    \label{fig:error-on-encryption-delta}
\end{figure}

CKKS employs various configuration parameters to guarantee a certain security level (e.g., 128 bits), such as the polynomial degree $N$, the ciphertext coefficient modulus $q$, and the scale factor $\Delta$. The scale factor is essential for adjusting the input data to maintain its precision as much as possible throughout the HE stages and, as discussed in this section, it influences error sensitivity. Figure~\ref{fig:error-on-encryption-delta} shows the $L_2$ norm error of the recovered output using different scale factors: $2^{20}$, $2^{40}$, and $2^{50}$. As observed, an increase in $\Delta$ leads to enhanced error resilience, characterized by a reduction in errors. The multiplication by the scale factor $\Delta$ effectively ``shifts right'' the coefficient, thereby augmenting the quantity of bits that remain unaffected by errors.

In practical applications, the scale factor $\Delta$ must be carefully adjusted for effective CKKS implementations. The choice of this factor represents a compromise between enhanced computation precision (larger $\Delta$) and reduced computational complexity (smaller $\Delta$). However, when incorporating error resilience into the analysis, such compromise needs to be revisited.

\subsection{RNS and NTT Optimizations Impact}

\begin{figure}[!ht]  
  \centering
  \begin{minipage}{0.28\columnwidth}  
    \centering
    \includegraphics[height=2.6cm]{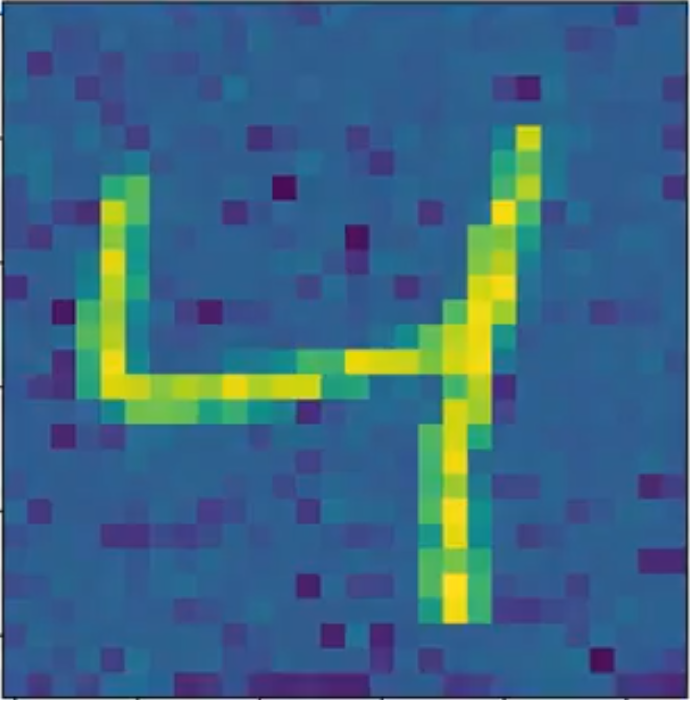}  % Ajusta la altura 
    \caption*{\small (a)~Original}  % Usar \caption* para evitar "Figura"
    \label{fig:decryptable}
  \end{minipage}\hfill
  \begin{minipage}{0.28\columnwidth}
    \centering
    \includegraphics[height=2.6cm]{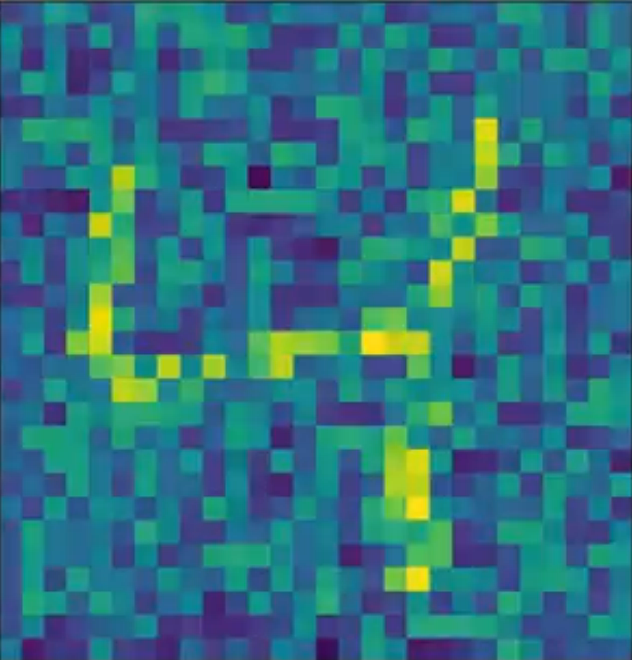}  % Ajusta la altura
    \caption*{\small (b)~No RNS/NTT}
    \label{fig:maybe}
  \end{minipage}\hfill
  \begin{minipage}{0.28\columnwidth}
    \centering
    \includegraphics[height=2.6cm]{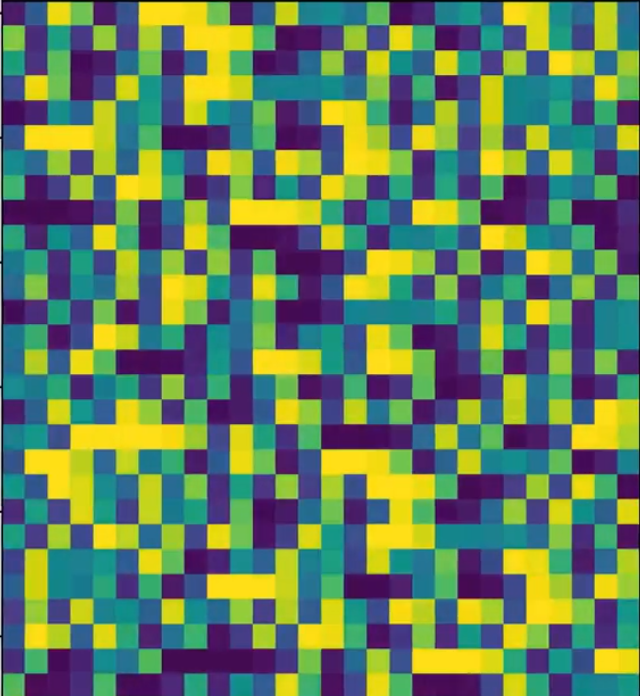}  % Ajusta la altura
    \caption*{\small (c)~RNS+NTT}
    \label{fig:undecryptable}
  \end{minipage}
  \caption{Impact of a single bit-flip in CKKS encoding.}
  \label{fig:ckks-error-example}
\end{figure}

Schemes like CKKS rely on two widely adopted optimizations to make HE problems tractable on available systems: the residue number system (RNS) and the number theoretic transform (NTT). RNS splits the huge polynomial coefficients used by HE schemes into smaller ones that fit common 64-bit processor words; while NTT enables efficient multiplication of high-degree polynomials. It is not the focus of this work to delve into the details of RNS and NTT (additional details can be found in~\cite{Cheon2019}). However, these optimization techniques have different effects on CKKS' error sensitivity. Figure~\ref{fig:ckks-error-example} illustrates a scenario in which an image from the MNIST dataset (Figure~\ref{fig:ckks-error-example}(a)) is processed through the HE pipeline. The effect of a bit error introduced during the encoding phase, without the application of RNS and NTT, is depicted in Figure~\ref{fig:ckks-error-example}(b), where a fairly accurate representation of the original image is retrieved. Conversely, when a bit error occurs during the utilization of RNS and NTT, the resulting image is entirely distorted, as demonstrated in Figure~\ref{fig:ckks-error-example}(c).

\begin{figure}[!ht]
    \centering
    \includegraphics[width=0.8\linewidth]{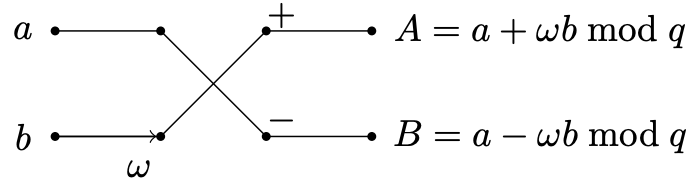}  
        \caption{Cooley-Tukey butterfly.}
    \label{fig:ct-butterfly}
\end{figure}

\noindent\textbf{RNS case:} RNS uses Equation~\ref{eq:crt} to reconstruct a polynomial coefficient from its smaller RNS remainders $r_k$. A bit error in $r_k$ is amplified when multiplied by the large $\left[\left(\frac{1}{Q_k}\right)\bmod q_k \right] Q_k$ factor, impairing the reconstruction of the original coefficient. 

\begin{equation}\label{eq:crt}
    p = \left(\sum_{k=1}^L r_k \left[\left(\frac{1}{Q_k}\right)\bmod q_k \right] Q_k \right) \text{ mod } Q
\end{equation}
\\
\noindent\textbf{NTT case:} The NTT, which is a variant of the Discrete Fourier Transform (DFT), can be implemented using Cooley-Tukey butterfly computations (Figure~\ref{fig:ct-butterfly}). As a result, a bit error in any of the NTT input elements ($a$ or $b$ in the figure) will inherently spread across \textit{all} the output elements ($A$ and $B$).

%\subsection{RNS and NTT Optimizations Impact}

%Figure~\ref{fig:rns-error} shows an example where a bit error is injected in the ciphertext $c = (c_0(X), c_1(X))$, affecting one of the $r_k$ reminders.

%\begin{figure}[!ht]
%    \centering
%    \includegraphics[width=0.8\linewidth]{figures/OpenFHE_Encrypt_CategorizecategoriesPlot_1_0_4_60_20_1_8_0_100_25.pdf}  
%        \caption{Level of decryption success of a single-bit flip in the ciphertext when RNS is enabled. The X-axis shows the position of the modified bit. Polynomial degree $N=8$.}
%    \label{fig:rns-error}
%\end{figure}

\section{Conclusion}

This work provides a first-cut analysis of error characterization in CKKS, a popular homomorphic encryption (HE) scheme due to its support of fixed-point arithmetic in AI and machine learning applications. The study examines the sensitivity of errors based on their occurrence location, the scale factor $\Delta$, and the implementation of residue number system (RNS) and number theoretic transform (NTT) techniques.

The findings reveal trade-offs between robustness, performance, and security that may be overlooked when robustness is not considered. For instance, while smaller scale factors are advantageous for minimizing HE complexity and enhancing the noise budget for operations, larger scale factors can enhance bit error resilience. Additionally, RNS and NTT techniques are acknowledged as significant optimizations in today's CKKS applications; however, their use may lead to severe consequences in the presence of bit errors.

The intrinsic error-centric nature of HE exacerbates this issue, as errors arising from defective hardware or software can easily blend with the inherent errors of the encryption, making detection difficult. Consequently, silent data corruption is anticipated to become a prevalent challenge in environments susceptible to faults in homomorphic encryption. Consequently, we anticipate that this research will lay the groundwork for significant future investigations in this domain.

%%%%%%
%% Appendix:
%% If needed a single appendix is created by
%%
%\appendix
%%
%% If several appendices are needed, then the command
%%
% \appendices
%%
%% in combination with further \section commands can be used.
%%%%%%

%\section*{Acknowledgment}

%You can add yor acknowledgment here.

%%%%%%
%% To balance the columns at the last page of the paper use this
%% command:
%%
%\enlargethispage{-1.2cm} 
%%
%% If the balancing should occur in the middle of the references, use
%% the following trigger:
%%
%\IEEEtriggeratref{7}
%%
%% which triggers a \newpage (i.e., new column) just before the given
%% reference number. Note that you need to adapt this if you modify
%% the paper.  The "triggered" command can be changed if desired:
%%
%\IEEEtriggercmd{\enlargethispage{-20cm}}
%%
%%%%%%

%%%%%%
%% References:
%% We recommend the usage of BibTeX:
%%
\bibliographystyle{IEEEtran}
\bibliography{refs}

\begin{thebibliography}{10}
\providecommand{\url}[1]{#1}
\csname url@samestyle\endcsname
\providecommand{\newblock}{\relax}
\providecommand{\bibinfo}[2]{#2}
\providecommand{\BIBentrySTDinterwordspacing}{\spaceskip=0pt\relax}
\providecommand{\BIBentryALTinterwordstretchfactor}{4}
\providecommand{\BIBentryALTinterwordspacing}{\spaceskip=\fontdimen2\font plus
\BIBentryALTinterwordstretchfactor\fontdimen3\font minus \fontdimen4\font\relax}
\providecommand{\BIBforeignlanguage}[2]{{%
\expandafter\ifx\csname l@#1\endcsname\relax
\typeout{** WARNING: IEEEtran.bst: No hyphenation pattern has been}%
\typeout{** loaded for the language `#1'. Using the pattern for}%
\typeout{** the default language instead.}%
\else
\language=\csname l@#1\endcsname
\fi
#2}}
\providecommand{\BIBdecl}{\relax}
\BIBdecl

\bibitem{Konstantinos2015}
K.~Konstantinos, M.~Persefoni, F.~Evangelia, M.~Christos, and N.~Mara, ``Cloud computing and economic growth,'' in \emph{Proceedings of the 19th Panhellenic Conference on Informatics}, 2015, pp. 209--214.

\bibitem{Sen2015}
J.~Sen, ``Security and privacy issues in cloud computing,'' in \emph{Cloud technology: concepts, methodologies, tools, and applications}.\hskip 1em plus 0.5em minus 0.4em\relax IGI global, 2015, pp. 1585--1630.

\bibitem{Williams1980}
H.~Williams, ``A modification of the {RSA} public-key encryption procedure (corresp.),'' \emph{IEEE Transactions on Information Theory}, vol.~26, no.~6, pp. 726--729, 1980.

\bibitem{Elgamal1986}
T.~Elgamal, ``A public key cryptosystem and a signature scheme based on discrete logarithms,'' \emph{IEEE Transactions on Information Theory}, vol.~31, no.~4, pp. 469--472, 1985.

\bibitem{Yi2014}
X.~Yi, R.~Paulet, E.~Bertino, X.~Yi, R.~Paulet, and E.~Bertino, \emph{Homomorphic encryption}.\hskip 1em plus 0.5em minus 0.4em\relax Springer, 2014.

\bibitem{Cheon2017}
J.~H. Cheon, A.~Kim, M.~Kim, and Y.~Song, ``Homomorphic encryption for arithmetic of approximate numbers,'' in \emph{Advances in Cryptology--ASIACRYPT 2017: 23rd International Conference on the Theory and Applications of Cryptology and Information Security, Hong Kong, China, December 3-7, 2017, Proceedings, Part I 23}.\hskip 1em plus 0.5em minus 0.4em\relax Springer, 2017, pp. 409--437.

\bibitem{Cheon2019}
J.~H. Cheon, K.~Han, A.~Kim, M.~Kim, and Y.~Song, ``A full {RNS} variant of approximate homomorphic encryption,'' in \emph{Selected Areas in Cryptography--SAC 2018: 25th International Conference, Calgary, AB, Canada, August 15--17, 2018, Revised Selected Papers 25}.\hskip 1em plus 0.5em minus 0.4em\relax Springer, 2019, pp. 347--368.

\bibitem{10.1145/1060590.1060603}
\BIBentryALTinterwordspacing
O.~Regev, ``On lattices, learning with errors, random linear codes, and cryptography,'' in \emph{Proceedings of the Thirty-Seventh Annual ACM Symposium on Theory of Computing}, ser. STOC '05, 2005, p. 84–93. [Online]. Available: \url{https://doi.org/10.1145/1060590.1060603}
\BIBentrySTDinterwordspacing

\bibitem{dixit2021silentdatacorruptionsscale}
\BIBentryALTinterwordspacing
H.~D. Dixit, S.~Pendharkar, M.~Beadon, C.~Mason, T.~Chakravarthy, B.~Muthiah, and S.~Sankar, ``Silent data corruptions at scale,'' 2021. [Online]. Available: \url{https://arxiv.org/abs/2102.11245}
\BIBentrySTDinterwordspacing

\bibitem{OpenFHE}
\BIBentryALTinterwordspacing
A.~A. Badawi, A.~Alexandru, J.~Bates, F.~Bergamaschi, D.~B. Cousins, S.~Erabelli, N.~Genise, S.~Halevi, H.~Hunt, A.~Kim, Y.~Lee, Z.~Liu, D.~Micciancio, C.~Pascoe, Y.~Polyakov, I.~Quah, S.~R.V., K.~Rohloff, J.~Saylor, D.~Suponitsky, M.~Triplett, V.~Vaikuntanathan, and V.~Zucca, ``{OpenFHE}: Open-source fully homomorphic encryption library,'' Cryptology ePrint Archive, Paper 2022/915, 2022, \url{https://eprint.iacr.org/2022/915}. [Online]. Available: \url{https://eprint.iacr.org/2022/915}
\BIBentrySTDinterwordspacing

\end{thebibliography}
%%
%% where we here have assumed the existence of the files
%% definitions.bib and bibliofile.bib.
%% BibTeX documentation can be obtained at:
%% http://www.ctan.org/tex-archive/biblio/bibtex/contrib/doc/
%%%%%%

\end{document}